\documentclass{acm_proc_article-sp}
\usepackage{graphicx}
\usepackage{url}
\usepackage[usenames]{color}
\usepackage{hyperref}
\usepackage{subfig}
\usepackage{caption}
\usepackage{hyphenat}
\DeclareCaptionType{copyrightbox}

\newcommand{\Msol}{{\rm M}_{\odot}}
\newcommand{\LCDM}{$\Lambda$CDM}

\newcommand{\OmDE}{\Omega_{\rm DE}}

\newcommand{\hinv}{\hbox{\, $h^{-1}$ }}

\newcommand{\gpc }{\hbox{\rm Gpc} }

\begin{document}

\conferenceinfo{XSEDE12}{July 16--20, 2012 Chicago, Illinois USA}

\title{A High Throughput Workflow Environment for Cosmological Simulations}

\numberofauthors{9}
\author{
\alignauthor
Brandon~M.~S.~Erickson\\
       \affaddr{Physics Department}\\
       \affaddr{University of Michigan}\\
       \affaddr{Ann Arbor, MI 48109}\\
       \email{bmse@umich.edu}
       \alignauthor
Raminderjeet~Singh\\
       \affaddr{Pervasive Technology Institute}\\
       \affaddr{Indiana University}\\
       \affaddr{Bloomington, IN 47408}\\
       \email{ramifnu@indiana.edu}
\alignauthor
August~E.~Evrard\\
       \affaddr{Physics Department}\\
       \affaddr{University of Michigan}\\
       \affaddr{Ann Arbor, MI 48109}\\
       \email{evrard@umich.edu}
       }

\additionalauthors{
Matthew R. Becker (Department of Physics, Kavli Institute for Cosmological Physics The University of Chicago, Chicago, IL 60637 email: {\texttt{beckermr@uchicago.edu}})
and
Michael T. Busha (Institute for Theoretical Physics, University of Z\"{u}rich, Z\"{u}rich, Switzerland and Physics Division, Lawrence Berkeley National Laboratory, Berkeley, CA 94720 email: {\texttt{mbusha@physik.uzh.ch}})
and
Andrey V. Kravtsov (Kavli Institute for Cosmological Physics, Department of Astronomy and Astrophysics, The University of Chicago, Chicago, IL 60637 email: {\texttt{andrey@oddjob.uchicago.edu}})
and
Suresh Marru (Pervasive Technology Institute, Indiana University, Bloomington, IN 47408 email: {\texttt{smarru@cs.indiana.edu}})
and
Marlon Pierce (Pervasive Technology Institute, Indiana University, Bloomington, IN 47408 email: {\texttt{mpierce@cs.indiana.edu}})
and
Risa H. Wechsler (Kavli Institute for Particle Astrophysics and Cosmology, Physics Department, Stanford University, Stanford, CA 94305 email: {\texttt{rwechsler@stanford.edu}})
}

\date{\today}

\maketitle


\begin{abstract}

The next generation of wide-area sky surveys offer the power to place extremely precise constraints on cosmological parameters and to test the source of cosmic acceleration.   These observational programs will employ multiple techniques based on a variety of statistical signatures of galaxies and large-scale structure.    These techniques have sources of systematic error that need to be understood at the percent-level in order to fully leverage the power of next-generation catalogs.  Simulations of large-scale structure provide the means to characterize these uncertainties.   We are using XSEDE resources to produce multiple synthetic sky surveys of galaxies and large-scale structure in support of science analysis for the Dark Energy Survey.  
In order to scale up our production to the level of fifty $10^{10}$-particle simulations, we are working to embed production control  within the Apache Airavata  workflow environment.  
We explain our methods and report how the workflow has reduced production time by 40\% compared to manual management.

\end{abstract}




\section{Introduction}\label{sec:intro}

A decade, and a Nobel Prize,\footnote{ \url{http://www.nobelprize.org/nobel_prizes/physics/} } after its discovery, the nature of cosmic acceleration remains a mystery.
Evidence continues to favor the simplest form of dark energy, so-called \LCDM\ models with a constant vacuum energy density (or cosmological constant) having a fixed ratio of pressure to energy density (equation of state parameter), $w = p/\rho = -1$ \cite{2009ApJS..180..306D, 2009ApJ...692.1060V}.  
Testing for departures from this canonical model requires large, sensitive astronomical surveys capable of delivering percent-level statistical constraints on $w$ and the dark energy density, $\OmDE$ \cite{2006astro.ph..9591A}.
Departures from \LCDM\ expectations may signal a time-varying equation of state parameter anticipated by specific theoretical models \cite{2008ARA&A..46..385F} or may indicate that gravity departs from general relativity on large scales \cite{2009PhRvD..80h3505S, 1998ARA&A..36..599B}.  

Realizing the full statistical power of upcoming surveys requires addressing all potential sources of systematic error associated with applying tests of cosmic acceleration based on the large-scale distribution of galaxies and clusters of galaxies.  
We are performing a suite of simulations that will allow us to address a range of sources of systematic error for the upcoming Dark Energy Survey (DES).\footnote{ \url{http://www.darkenergysurvey.org} }
The simulations can also be used to improve theoretical calibration of the clustered matter distribution, including the abundance and clustering of massive halos \footnote{The term halos refers to self-bound, quasi-equilibirium structures that emerge via gravitational collapse of initial density peaks.} that host galactic systems.  The particular set of simulations we are performing on XSEDE in 2012 will form the basis of a Blind Cosmology Challenge for the DES collaboration.  




\section{Cosmological simulations for DES}\label{sec:simulation}


The DES \cite{2005astro.ph.10346T, 2005astro.ph.10194A, 2005astro.ph.10195A} is a Stage III\footnote{In the language of the Dark Energy Task Force, see \cite{2006astro.ph..9591A}} dark energy project jointly sponsored by DoE and NSF that is on track to see first light in the fall of 2012. 
The project will use a new panoramic camera on the Blanco 4-m telescope at the Cerro Tololo Inter-American Observatory in Chile to image $\approx 5000$ square degrees of the sky in the South Galactic Cap in four optical bands, and to carry out repeat imaging over a smaller area to identify distant type Ia supernovae and measure their lightcurves.
In addition, the main imaging area of the DES overlaps the South Pole Telescope\footnote{ \url{http://pole.uchicago.edu/} } sub-mm survey that will identify galaxy clusters via the Sunyaev-Zel'dovich effect as well as the VISTA\footnote{ \url{ http://www.vista.ac.uk} } infrared survey of galaxies, which will provide additional information on galaxy photometric redshifts and on the properties of galaxy clusters at large cosmological redshift, $z > 1$.\footnote{A redshift, $z$, derived from spectroscopy measures both distance and look-back time to the source.  A galaxy at $z = 1$ emitted its light when the universe was $6.2$ Gyr old and it lies at a distance of $3.3$~Giga-parsecs from the Milky Way.}  Roughly three hundred scientists across nearly thirty institutions comprise the DES collaboration.

The DES will be the first project to combine four different methods to probe the properties of the dark sector (dark matter and dark energy) and test General Relativity gravity via evolution of the Hubble expansion parameter and the linear growth rate of structure.  The methods---baryon acoustic oscillations in the matter power spectrum, the abundance and spatial distribution of galaxy groups and clusters, weak gravitational lensing by large-scale structure, and type Ia supernovae---are quasi-independent.  
Each has sources of systematic error associated with it, some of which are unique to the method but many of which are shared.  Examples of the latter are the accuracy of photometric redshift estimates,\footnote{Photometric redshifts are distance estimates that use the multi-color fluxes measured in broad optical-IR bands as essentially a (very) low-resolution galaxy spectrum.} the form of the non-linear matter clustering power spectrum, and shape measurement errors for galaxy images that affect cosmic shear and galaxy cluster mass estimates.  DES will thus be the first survey to address joint systematics in multiple methods probing accelerating expansion of the universe.  N-body simulations provide key support for the analysis of systematics in the three methods associated with cosmic large-scale structure (all but supernovae in the above list).  To validate science analysis codes, the DES Simulation Working Group is coordinating a Blind Cosmology Challenge (BCC) process, in which a variety of sky realizations in different cosmologies will be analyzed, in a blind manner, by DES science teams.

\subsection{Blind Cosmology Challenge}\label{ssec:bcc}

The Blind Cosmology Challenge (BCC) process will require generating multiple galaxy catalogs to the full photometric depth across the full 5000 square degrees of the DES survey.   The effort will require roughly 6M SUs and generate 300 TB of output.   

Competing effects drive our simulation requirements.   On one hand, the dark matter distribution needs to be modeled within a large cosmic volume.
On the other hand, galaxy surveys also sample the nearby population of dwarf galaxies, implying a need for high spatial and mass resolution.  We have developed an approach that generates a set of discrete N-body sky realizations of dark matter structure spanning a range in resolution and volume.  We then dress this dark matter distribution with galaxies brighter than the DES limiting magnitudes in each passband.  

To implement the BCC, we plan to produce fifty $10^{10}$-particle N-body runs on XSEDE resources over the next two years.  
The collaboration has requested that BCC models explore a variety of cosmologies, parameters of which are known only to the Simulation Working Group members.  After a blind processing period, constraints on cosmological parameters from the science teams will be compared against their true values, gauging the validity of the processing pipelines.
We detail ongoing and proposed simulations along with our strategy for producing synthetic sky surveys of sufficient area and depth in the next section.

\subsection{Computational requirements}\label{ssec:compreq}

\begin{figure}[htbp]
 \centering 
\includegraphics[width = 0.75\linewidth]{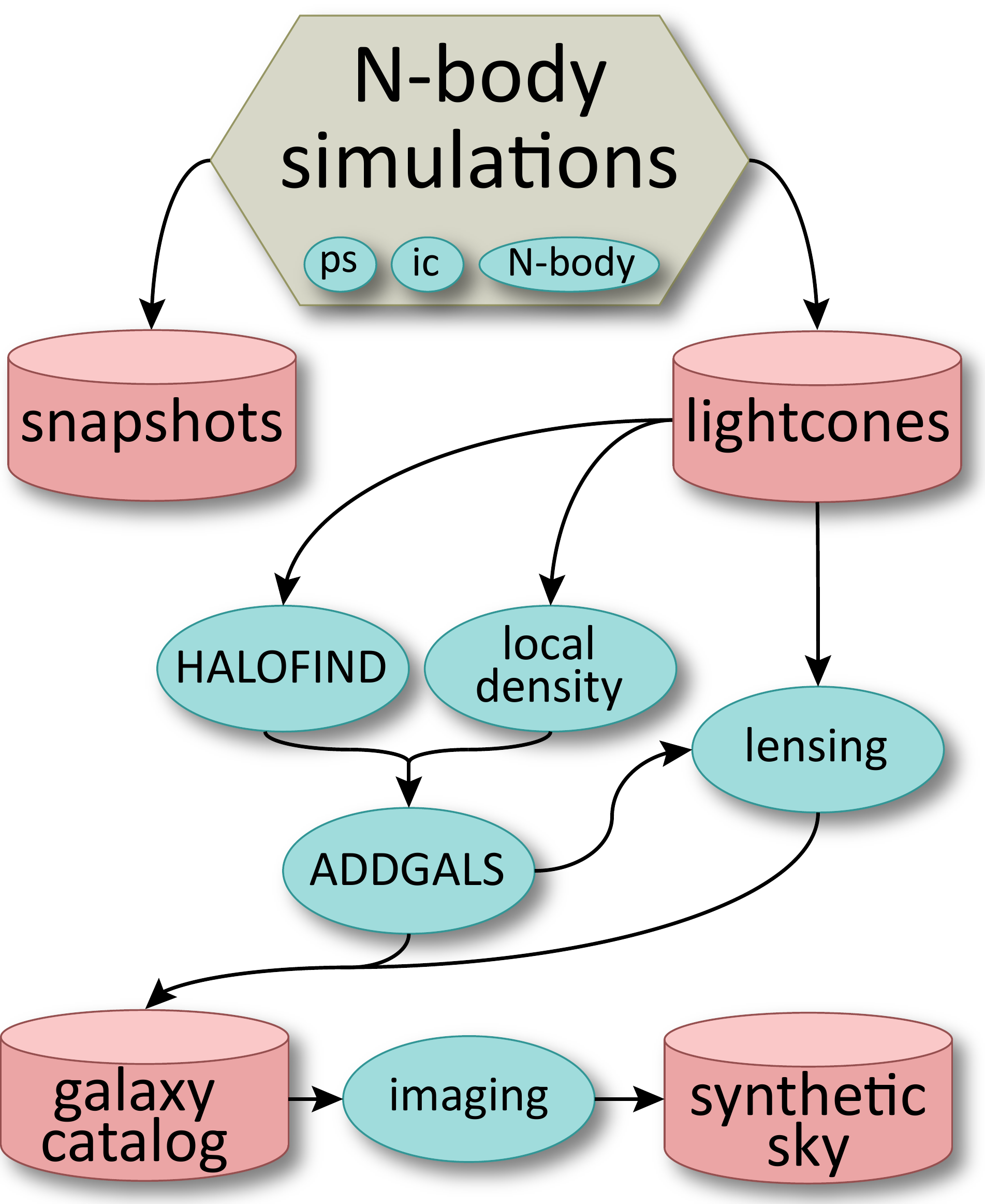}
  \caption{Processing steps to build a synthetic galaxy catalog are illustrated here and described in the text.  The XBaya workflow currently controls the top-most element (N-body simulations) which consists of methods to sample a cosmological power spectrum (ps), generating an initial set of particles (ic) and evolving the particles forward in time with Gadget (N-body).  The remaining methods are run manually on distributed resources. }
\label{fig:cosma}
\end{figure}

The dark matter structure from N-body simulations of a given cosmology forms the basis for galaxy catalog expectations.
The N-body models we run on XSEDE resources store particle configurations, $\{ x_i(t), v_i(t) \}$, in two forms: {\it snapshots} of the positions and velocities of all particles $i$ in the simulation volume at a fixed time $t$, and {\it lightcones} \cite{evrard02} that hold kinematic information for particles lying on the past lightcone of a virtual observer located at a fixed position, $x_{\rm cen}$, in the computational volume.\footnote{These particles satisfy the combined space-time requirement, $| x_{i}(t_{i}) - x_{\rm cen} | = r_{\rm cosm}(t_i)$, where $r_{\rm cosm}(t)$ is the cosmic metric distance as a function of time, a known function of the cosmological parameters.}
Once the N-body steps are completed, we proceed with additional processing using a combination of resources, including XSEDE, SLAC and other collaboration institutions.  The post-processing aims to create a synthetic DES catalog of galaxy properties derived from the N-body lightcone output.  Figure~\ref{fig:cosma} shows a schematic representation of the post-processing workflow.  

Halos, bound systems that host galaxies and clusters of galaxies, are identified in these outputs and their properties can be used to determine their central galaxy characteristics.  Halos, as well as local density estimates in under-resolved locations, are used by the ADDGALS\footnote{for Adding Density-Determined GAlaxies to Lightcone Simulations} algorithm to assign galaxy properties to suitably selected dark matter particles.  The matter along the past lightcone 
also sets the gravitational lensing shear signal applied to these galaxies, and we are developing a new multi-grid, spherical harmonic algorithm for this computation.  
Finally, galaxy catalogs that include lensing shear signals are processed with telescope/instrument/noise effects to produce images expected from the back-end electronics of the Dark Energy Camera.   From these images, which are generated for only 200 sq deg of sky, we can develop effective transfer functions to create full 5000 sq deg synthetic DES catalogs that contain realistic errors such as star/galaxy mis-classification, blended sources, and appropriate photometric errors.

Generating a cosmological N-body simulation has three main steps, described further below.  The first two create an initial particle set consistent with the structure expected at an early time of a chosen cosmological model.  The first step samples the linear perturbation spectrum of the cosmological model while the second step realizes that density and velocity field with a set of particles. 
For the latter, we use a second-order Lagrangian perturbation theory code (2LPTic) code that has been robustly tested in the community \cite{2006PhRvD..73f3519C, 2006MNRAS.373..369C}.
The amount of CPU time required to make these initial conditions is small (typically 300-400 CPU hours for simulations of the scale described here), but the memory requirements are more substantial (slightly less than 2~GB per core), although easily achievable on XSEDE machines.

The final step evolves this particle distribution under its self-gravity within an expanding cosmological background.   For this purpose, 
we are using a lean version of the Gadget cosmological N-body code \cite{2005MNRAS.364.1105S, 2005Natur.435..629S} modified by us to generate output along the past lightcone of synthetic observers in the computational volume.
We have worked on XSEDE resources for the past year to modify and optimize the code.  
The lean version has significantly reduced memory overhead, 44 bytes per particle compared to 84 for standard Gadget.
This reduction allows the simulations to fit on a smaller number of processors, affording better scaling.
These simulations require 50--100k CPU hours depending on the parameters and resolution of the computation, and generate up to 10~TB of total data output.


\section{Workflow Abstractions}\label{sec:workflow}
The simulation codes discussed in Section \ref{sec:simulation} are executed on large scale XSEDE resources managed by batch resource managers. The heterogeneity and complexity in interfacing with these resources slow down the computational scientists in harnessing the vast amount of available computing power. The eScience workflow systems abstract out these complexities and enable the use of innovations made in computational middleware. Scientific workflows are one of the prominent abstractions that allow scientist the carry out their scientific discovery and experimentation without having to worry about the underlying complexity. These abstractions, while lowering the entry and learning curves, also become more relevant to address human inefficiency to monitor long running jobs.   

To build our cosmological workflow, we leverage the experience and software developed by the Open Gateways Computing Environments project~\cite{ogcePierceTg10} facilitated by the XSEDE Extended Collaborative Support Services. The workflow infrastructure is based upon the Apache Airavata~\cite{airavataMarruGCE2011} framework. We are going to briefly describe the integration of the simulation codes with the workflow infrastructure and specific customizations made to the framework itself. Further details about the framework and its comparison to other workflow solutions are discussed in~\cite{airavataMarruGCE2011}. 

The Airavata workflow system is primarily targeted to support long running scientific applications on computational resources. Airavata's XBaya is a graphical workflow tool, allows composition, execution and monitoring of the workflows. The Airavata workflow engine requires these applications to be raised to a common abstraction that can be accessed using a standard protocol. The Airavata Generic Application Factory (GFac) component bridges this gap between applications and the workflow systems by providing a network accessible web service interface to the scientific application. 

\subsection{Implementation}\label{ssec:implem}

Once the simulation codes are deployed on XSEDE computational resources, we register descriptions of these applications with the Apache Airavata registry service. These descriptions are used by the Airavata GFac component to generate the artifacts required to expose the application as a service. The workflow developer can access these wrapped application services and construct workflows and orchestrate executions on target compute resources. The resulting workflow abstractions reduce human inefficiencies by providing a uniform interface for the scientist and hiding unnecessary complexities. 

To illustrate the construction of a cosmological workflow, we will describe the process in developing the N-body simulation workflow of the process illustrated in  Figure~\ref{fig:cosma}. Firstly the nature of the applications, its execution characteristics, and its input and output data are analyzed. The application meta information, including the executable location, its nature like serial or MPI, inputs and outputs, are described and registered with Airavata registry.  This process was followed for the following four applications. 

\begin{figure*}[t]
   \centering
   \includegraphics[width=\textwidth]{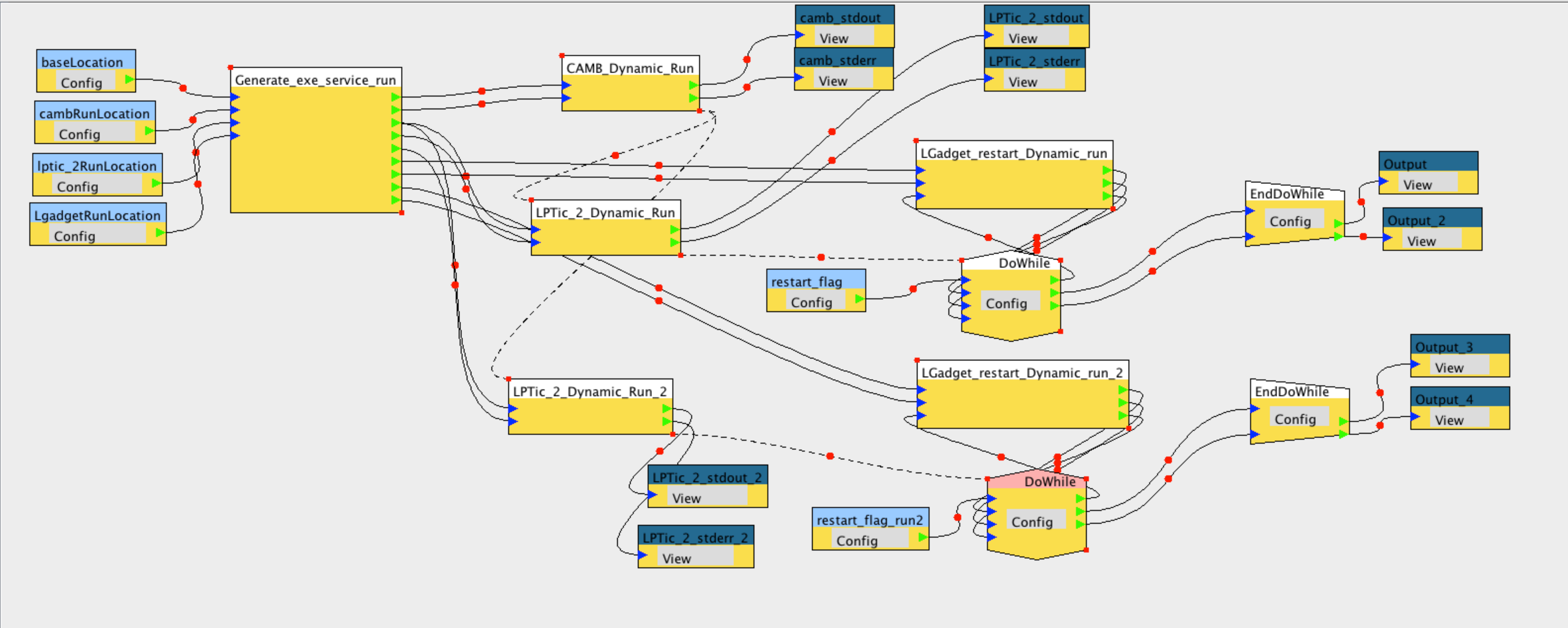} 
   \caption{XBaya workflow for a full cosmology with automatic queue resubmission if wallclock limits are reached and on-the-fly executable compilation.}
   \label{fig:desworkflow}
\end{figure*}

\begin{description}
\item[BCC Parameter Maker] \hfill \\
This initial setup code is written as a {\sc python} script and prepares necessary configurations and parameter files for the workflow execution. This simple script is forked on the XSEDE Ranger job management nodes. 
\item[CAMB] \hfill \\ 
The CAMB (Code for Anisotropies in the Microwave Background) \cite{2000ApJ...538..473L} application computes the power spectrum of dark matter, which is necessary for generating the simulation initial conditions. This application is as a serial {\sc fortran} code. The output files are relatively small ASCII files describing the power spectrum.
\item[2LPTic] \hfill \\
The Second-order Lagrangian Perturbation Theory initial conditions code \cite{2006MNRAS.373..369C} (2LPTic) is programmed using Message Passing MPI {\sc C} code that computes the initial conditions for the simulation from parameters and an input power spectrum generated by CAMB.  The output of this application are a set of binary files that vary in size from $\sim$80--250~GB depending on the simulation resolution. 
\item[LGadget] \hfill \\
The LGadget simulation code is MPI based {\sc C} code that uses a TreePM algorithm to evolve a gravitational N-body system \cite{2005MNRAS.364.1105S, 2005Natur.435..629S}.  The outputs of this step are system state snapshot files, as well as lightcone files, and some properties of the matter distribution, including diagnostics such as total system energies and momenta. The total output from LGadget depends on resolution and the number of system snapshots stored, and approaches close to 10 TeraBytes for large DES simulation volumes.
\end{description}

After all the above applications are registered, the Blind Cosmology Challenge Workflow is constructed using Airavata XBaya.  The resultant workflow graph is shown in  Figure~\ref{fig:desworkflow}. The workflow provides capabilities to configure the generation of initial conditions as well as the full N-body simulation components of the BCC process.  

\subsection{Workflow System Enhancements}\label{ssec:workflowEnhancemnets}

\textbf{Iterative execution support for long running applications}
The N-Body simulation requires multiple days of execution, but the XSEDE Ranger cluster limits maximum wall time of 48 hours. To mitigate this limitation, the workflow infrastructure has to allow iterative support so the job can be broken down into multiple increments of 48 hour jobs harnessing the check-point restart capabilities within the application. These capabilities required sophistication beyond the blind restarts, in order to account for application execution patterns and exception handling. These capabilities can be potentially matured into a formal Do-While construct semantics of workflow engines.  

\textbf{Output Transfers}
The workflow executions tend to produce terabytes of data residing on the cluster scratch file systems and have to be persisted for a longer durations. The data movement to archival systems like TACC Ranch for long term storage have to be provided. The large file data movement is non-trivial process. Even though many advancements have been made in this area, the seamless reliable data transfers are still challenging. The emerging solutions like Globus Online~\cite{globus-online}, GridFTP client API\cite{GridFTP-API} and bbcp~\cite{bbcp} are potentially viable options. We leverage the Ranch archival system mounting on Ranger and use bbcp to copy the workflow outputs for test runs. We have yet to explore this as a solution for production executions. To transfer data to the post processing remote locations, Globus Online and GridFTP client are viable options. 



\section{Results}\label{sec:res}

Toward our goal of fifty cosmological simulations over two years, we have completed seven on XSEDE resources in the final quarter of  2011 and the first quarter of  2012.  Most of these simulations were run `by hand,' while the last was performed using the XBaya workflow environment \S \ref{sec:workflow}.  We summarize the completed simulations in Table \ref{tab:cosmoRuns}.

For a given cosmology, we generate four N-body simulations in nested volumes, consisting of three large-volume realizations with $2048^3$ particles and one smaller volume of $1400^3$ particles.   This approach allows a better match to halo mass selection imposed by the magnitude-limited nature of the DES galaxy sample.  As indicated in Table \ref{tab:cosmoRuns}, the mass resolution varies by nearly a factor of 60 from our smallest to largest volumes.  A halo resolved by a minimum of 100 particles ranges from a mass of  $3 \times 10^{12} \hinv \Msol$ in the near-field simulation to $2 \times 10^{14} \hinv \Msol$ in the far-field.\footnote{Here, $h$ denotes Hubble's constant, ${\rm H}_0$, in dimensionless form, $h = H_0 /100 \, {\rm km \, s}^{-1} \, {\rm Mpc}^{-1}$.}  The former is roughly the mass of our Milky Way galaxy's halo while the latter corresponds to the mass scale of clusters of galaxies.

\begin{figure*}[t]
   \centering
\includegraphics[width = \textwidth]{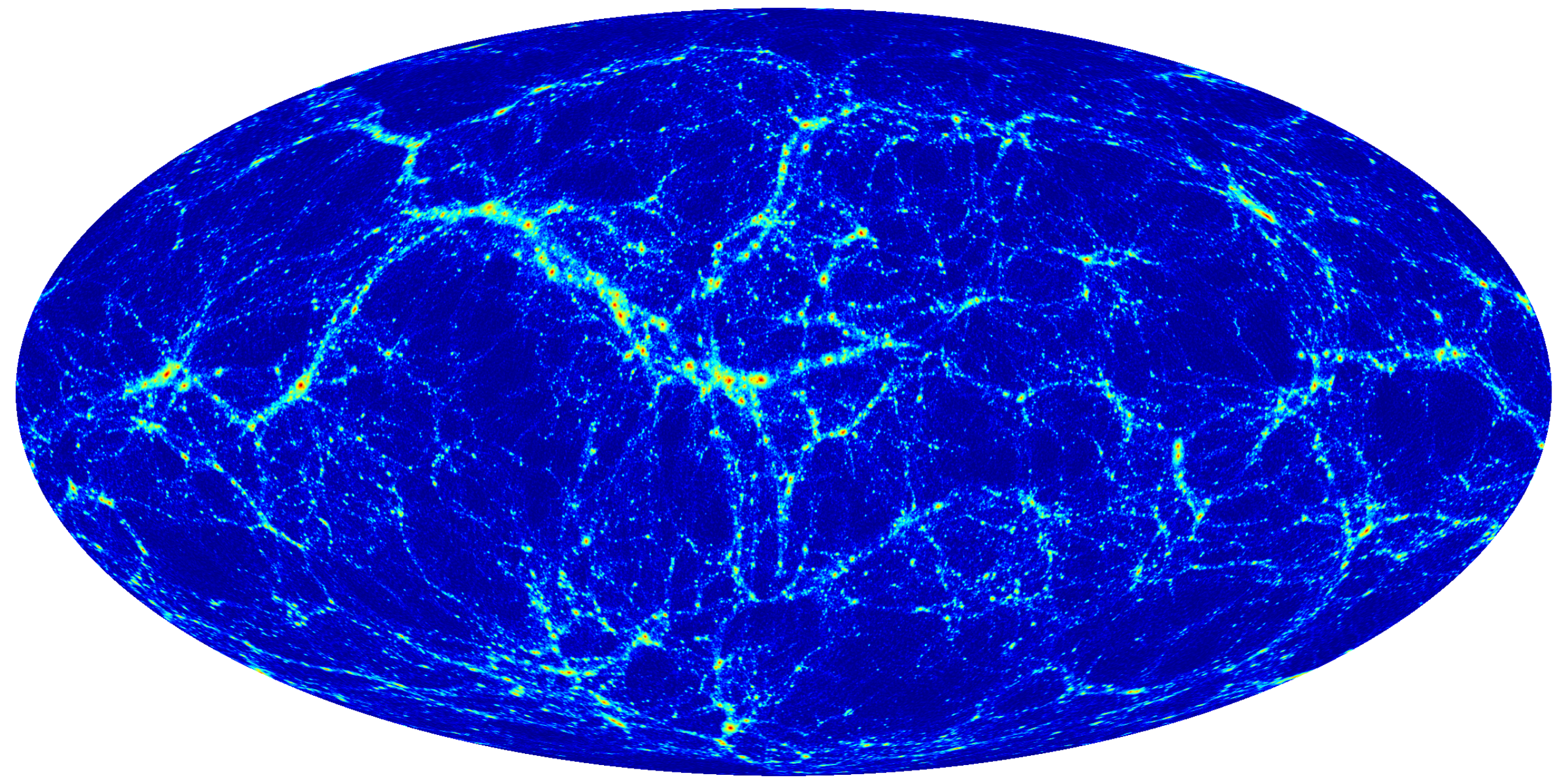}
  \caption{Full-sky image of the dark matter  density in a thin radial slice (50--75$\hinv$Mpc distance from observer) taken from our 1050$\hinv$Mpc \LCDM\ simulation.  Color maps the local matter density relative to the mean value on a logarithmic scale ranging from -1 (blue) to roughly 500 (red).  }
\label{fig:LCimage}
\end{figure*}

Each simulation produces lightcone outputs centered on each of the eight corners of the computational volume.  By employing the periodic boundary conditions of the computational domain, we can stitch these octants of sky into a single $4\pi$ representation of the full past lightcone of a hypothetical observer placed at the origin of the simulation.  A map of the resultant structure in a thin radial slice of synthetic sky is shown in Figure~\ref{fig:LCimage}.   Along with these lightcone outputs, we also record snapshots of the particle configuration in the full volume at 20 epochs, leading to an overall data output of $8.4$~TB for the $2048^3$-particle runs.  

\begin{table}
\centering
\caption{Summary of completed simulations, giving the side length $L$ (in $\hinv \gpc$) of the periodic cube, number of particles, particle mass (in $10^{12} \hinv \Msol$), number of simulations run, kiloSUs used for the run as well as any completed postproccessing and the amount of data generated in Terabytes.}
\begin{tabular}{ l  c  c  c  c  c  c }
$ L$ & $N_{\rm part}$ & $M_{\rm part}$ & $N_{\rm run}$ & kSU & Data (TB)  \\
\hline
1.05 & $1400^3$ & $0.027$ & 2 & 121 & 5.4 \\
2.60 & $2048^3$ & $0.131$ & 2 & 284 & 16.8 \\ 
4.00 & $2048^3$ & $0.476$ & 2 & 149 & 16.8 \\
6.00 & $2048^3$ & $1.650$ & 1 & 95 & 8.4 \\ 
 \hline
All & $-$ & $-$ & 7 & 649  & 47.4 \\
\end{tabular}
\label{tab:cosmoRuns}
\end{table}

The combined lightcone files for a single cosmology, $\sim$2 TB of data, are transferred manually via Globus Online to SLAC for the post-processing steps, illustrated in Figure~\ref{fig:cosma}, that create DES galaxy catalogs.
We first identify dark matter halos using a new algorithm, dubbed ROCKSTAR 
\cite{2011arXiv1110.4372B}, that uses a direct socket-to-socket task-scheduling approach to operate efficiently on sub-regions of large simulations.
We determine a local Lagrangian density by computing the distance to the Nth nearest particle, where N is chosen to enclose a mass of $\sim$$10^{13}\, \Msol$, roughly the transition mass above which halos host more than one bright galaxy.
With the dark matter halos and Lagrangian density estimate in place, the ADDGALS algorithm creates a synthetic galaxy catalog for science analysis.  Finally, gravitational lensing shear is computed from the lightcone matter distribution, and its effect on galaxy images recorded.   A single post-processed galaxy catalog, with 102 parameters per galaxy, is an 0.5 TB dataset.

\subsection{Efficiency gains with XBaya}\label{ssec:eff}

The XBaya workflow shown in Figure \ref{fig:desworkflow} was tested and refined using smaller simulations  over the period Oct 2011 to Mar 2012.  We transitioned to production use for our most recent simulations.   One workflow-managed simulation has run to completion, while a second job crashed because of a hardware problem on TACC Ranger.  

Even with this limited information, we can compare the efficiency of running the required jobs under XBaya to those previously run manually.   Jobs were submitted to the long queue at TACC Ranger, which has maximum resource limits of 1024 processors and 48 hour runtime.  In Table \ref{tab:efficiency}, Total Time is the wallclock time interval for the entire production process, while CPU Time gives the sum of the run times of the required jobs.   Times reflect the full N-body production process, from generating initial conditions all the way through to completing of the final N-body timestep.   Efficiency is the ratio of CPU to Total times, with $100\%$ representing the ideal scenario of running without interruption.

\begin{table}[t]
\centering
\caption{Comparison of Manual and Workflow-enabled production times}
\begin{tabular}{l l l l}
Run & Total Time & CPU Time & Efficiency \\ \hline
Manual & 8:15:33:05 & 4:07:24:10 & 50.0\% \\
Manual & 4:05:39:07 & 2:17:50:06 & 64.8\% \\
Workflow & 2:09:53:23 & 2:05:28:09 & 92.4\% \\
\end{tabular}
\label{tab:efficiency}
\end{table}%

The first two rows of In Table \ref{tab:efficiency} list different manually-processed simulations.  The first is a large simulation that needed four total submissions to the queue: one for initial conditions, and three for N-body  computation.   The second row was a smaller job requiring three submission, one for initial conditions and two for N-body.   These runs are relatively inefficient because, each time the wallclock limit is reached, the user is notified via email, and then must log back into the cluster to submit the next job.
If the wallclock limit is reached at an inconvenient time, considerable time can elapse before the next submission.  More submissions tends to drive up the inefficiency.

The last row shows the efficiency of running the full production process via the XBaya workflow environment.   By enabling immediate submission of jobs when the preceding job finishes, the efficiency improves to 92\%, well above the 50--60\% found for manual processing.   Under the workflow, the only time spent not computing is spent waiting in the queue time on the cluster.   That is, our job production becomes limited only by the instantaneous compute resources available on TACC Ranger.  

We found that the workflow can also help prevent errors in simulation set-up.  Our first production level workflow was designed to have the same parameters as a pair of manually completed simulations.  We soon found that the latter, `by hand' simulations were inconsistent with the workflow simulations.  Investigating the source of the inconsistency, we found that the workflow was correct, and that a parameter had been mistakenly set to a wrong value in the original simulations.   This shows the workflow's value in reducing the risk of simple human run-time errors.

\subsection{DES science}\label{ssec:science}

Synthetic surveys from the first set of simulations are currently in use by DES science groups.  We provide here two examples of analyses from cluster and weak lensing science groups.  

Members of the Galaxy Cluster Working Group are using the synthetic galaxy catalogs to evaluate different methods for identifying the massive halos that host clusters.   Because 5-band optical photometry provides relatively crude distance information for each single galaxy, the ability to identify localized spatial clusters is compromised by poor depth resolution.  Different cluster finding algorithms have methods to mitigate this loss of information, but none has been applied to a large survey with the depth of DES.   

Model galaxies are assigned to specific, unique dark matter halos, so an ideal cluster finder would return the list of original halos with all members intact.  In practice, the distance errors, incorrect choice of cluster location and other effects imply that the set of clusters found does not perfectly match the input set of halos.  However, a correspondence map between cluster and halo sets can be derived using joint galaxy membership criteria.   The utility of a cluster finder can then be characterized by two parameters: i) purity, the fraction of clusters that correspond to genuinely massive halos, and; ii) completeness, the fraction of halos that are found by the cluster finder.   The ideal value for both parameters is one. 

\begin{figure}[t]
 \centering 
 \subfloat[Purity]{\label{fig:purity}\includegraphics[width=0.8\linewidth,height=0.75\linewidth]{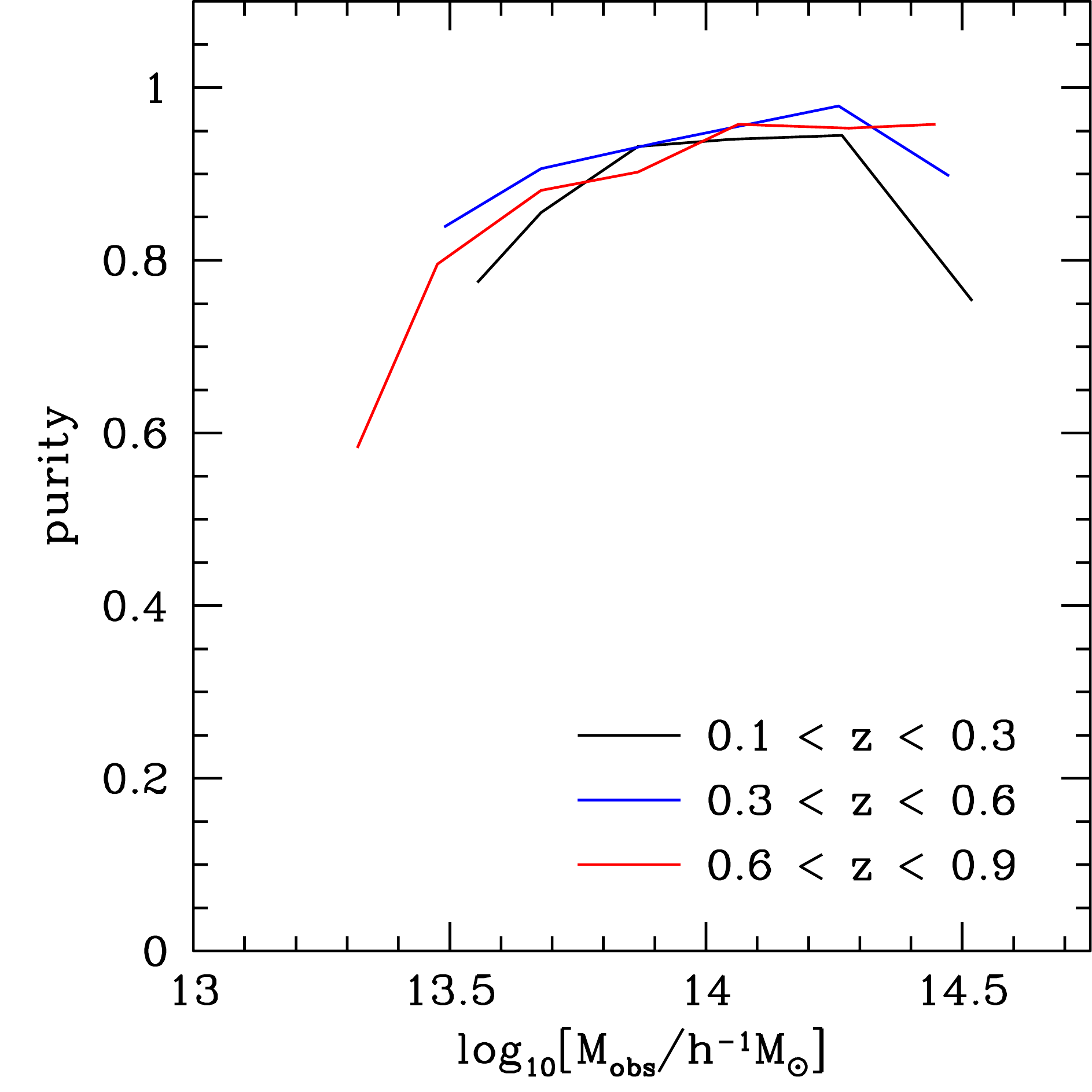}}\\
 \subfloat[Completeness]{\label{fig:completeness}\includegraphics[width=0.8\linewidth,height=0.75\linewidth]{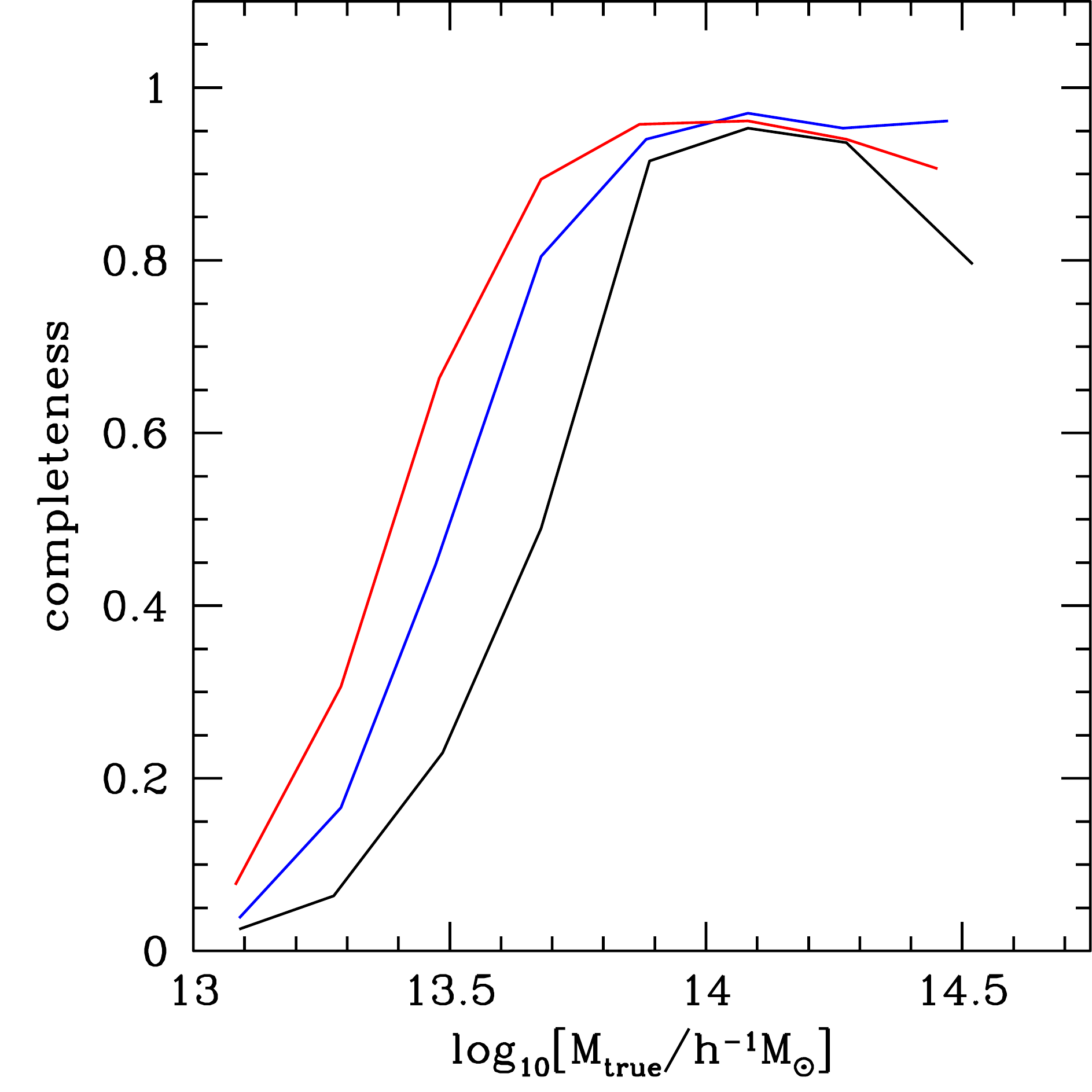}}
  \caption{Purity as a function of the observed mass proxy (top) and completeness as a function of true halo mass (bottom) of the redMaPPer galaxy cluster finding algorithm applied to the first synthetic DES sky catalog.  Colors denote the redshift intervals shown in the legend. }
\label{fig:purcomp}
\end{figure}

Figure \ref{fig:purcomp} shows recent purity and completeness measurements from the redMaPPer cluster finder \cite{2012ApJ...746..178R}.
The redMaPPer cluster finder works on so-called red-sequence galaxies, galaxies with evolved stellar populations that tend to be found in massive halos.  The algorithm is a Bayesian method that assigns galaxies a cluster membership probability based on the galaxy's color, and the density of nearby galaxies of a similar color.  Clusters are identified as clumps of similar high-likelihood member galaxies.   Figure~\ref{fig:purcomp} shows that redMaPPer is $> 80\%$ pure and has similarly high completeness above a redshift-dependent minimum halo mass.

\begin{figure}[htbp]
 \centering 
 \includegraphics[width = 0.9 \linewidth]{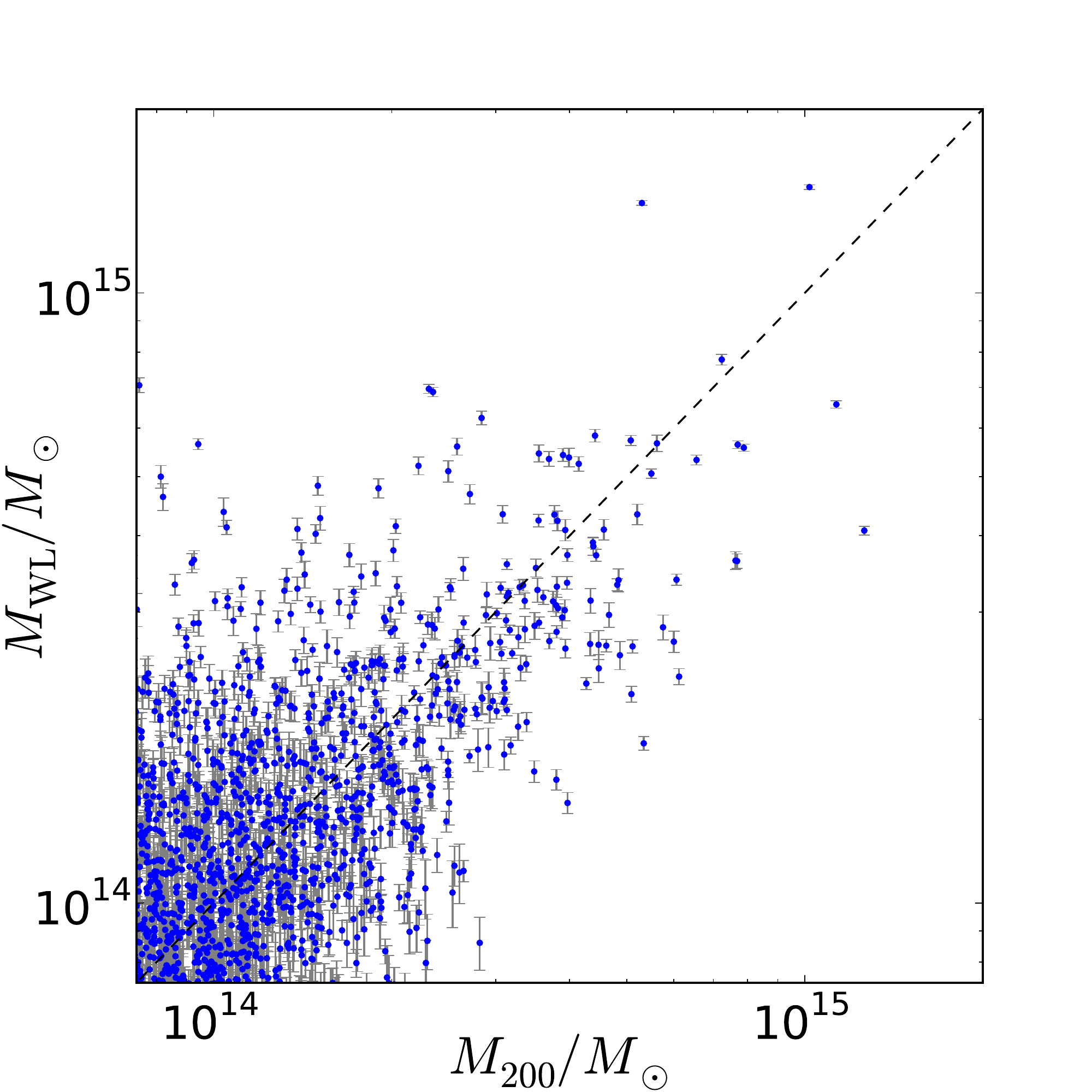}
  \caption{Comparison of the mass inferred from weak lensing shear analysis ($y$-axis) to the true halo mass ($x$-axis) for several thousand galaxy clusters identified in the first synthetic DES sky survey.  The dotted line is the identity relation.}
\label{fig:wlcalib}
\end{figure}

In a related project, members of the Weak Lensing Working Group have been investigating how best to estimate the mass of an observed cluster based on the weak lensing distortion of background galaxies.
Light from faint background galaxies is bent by the gravitational field of a massive halo, distorting the shape (shear) and density (magnification) of the galaxies.
Using a maximum likelihood estimator \cite{2000A&A...353...41S} to fit the synthetic  observations, and comparing to the true halo masses from the simulation, the group can understand how best to estimate masses and also provide feedback on the model used to generate the synthetic observations.  A recent calibration plot comparing estimated and true masses is shown in  Figure~\ref{fig:wlcalib}.

\subsection{Future Directions}\label{ssec:future}

Implementing the Airavata workflow for this project has entailed some overhead.  Scripts that set up the input parameter files needed to be developed, and new features were added to the existing codebase so that our applications would integrate more effectively with the workflow framework.  Interaction between the co-authors of this document---domain scientists along with TeraGrid AUSS team members---was essential to achieving a production-level service.   The effort invested has been worthwhile, with a significant gain in realized efficiency (Table~\ref{tab:efficiency}) for our first production run.  

In the near term, we have applied for continued XSEDE ECS support aimed at integrating the postprocessing and catalog production stages into the Airavata workflow (see  Figure~\ref{fig:cosma}).   We also want to integrate data movement into the workflow.  In addition, we have requested time at two XSEDE facilities (TACC Ranger and UCSD Trestles), and would like to expand the workflow logic to choose execution location in real time, based on queue loads.  

We also would like to use Airavata to improve our provenance practices.  The workflow system can capture provenance, including information such as when the data set was created, by whom, where, and with what application version and which input parameters.  Improved provenance can enable broader forms of sharing, reuse, and long-term preservation of our simulations and the resultant galaxy catalogs.  Additional development could include standardizing an API for simulation parameter input and output, so that other codes could be easily implemented in the workflow, such as N-body models that employ modified gravity \cite{2008PhRvD..78l3523O, 2012JCAP...01..051L}.

In the longer term, we could also expand our scope, generalizing our galaxy catalog construction process into a science gateway that would support broader classes of astrophysical studies.   The optical catalogs we create could be augmented by synthetic surveys at other wavelengths, from radio to X-ray, and our focus on galaxies could be expanded to include quasars, galactic stars, and other astrophysical objects.




\section{Summary}\label{sec:summ}

To meet the challenge of interpreting Big Astronomical Data for cosmological and astrophysical knowledge, new modes of study that incorporate model expectations derived from sophisticated simulations will be required.  Growing demand for simulated data products motivates the automation of simulation production methods within grid-aware workflow environments.  This work represents a first step in that direction.  

We are using XSEDE resources to produce multiple synthetic sky surveys of galaxies and large-scale structure in support of science analysis for the Dark Energy Survey.  To scale up our production to the level of fifty $10^{10}$-particle simulations, we have embedded production control within the Apache Airavata  workflow environment, resulting in a significant increase in production efficiency compared to manual management.






\section{Acknowledgments}
This work is partially supported by the Extreme Science and Engineering Discovery Environment (XSEDE) Extended Collaborative Support Service (ECSS) Program funded by the National Science Foundation through the award: OCI-1053575, and by 
NSF OCI Award 1032742, SDCI NMI Improvement: Open Gateway Computing Environments---Tools for Cyberinfrastructure\hyp Enabled Science and Education.   AEE acknowledges support from NSF AST-0708150 and NASA NNX07AN58G.
We thank Jeeseon Song, J\"{o}rg Dietrich, Eli Rykoff, and Eduardo Rozo for providing assistance with Figures 4 and 5.


\bibliographystyle{abbrvurl}
\bibliography{xsedeetal}

\balancecolumns
\end{document}